\documentclass[12pt]{article}

\usepackage[superscript]{cite}

\usepackage{times}
\usepackage{amssymb}
\usepackage{amsmath}
\usepackage{graphicx}
\usepackage{bm}
\usepackage{dcolumn}
\usepackage{color}
\usepackage{ulem}
\usepackage{float}
\usepackage{textcomp}
\restylefloat{table}
\usepackage{verbatim}
\usepackage{epstopdf}
\usepackage[colorlinks]{hyperref}
\usepackage[mediumspace,mediumqspace,squaren]{SIunits}

\newenvironment{sciabstract}{%
\begin{quote} \bf}
{\end{quote}}

\title{Bloch Ferromagnetism of Composite Fermions}

\author{Md. Shafayat Hossain,$^{1, \dagger}$ Tongzhou Zhao,$^{2, \dagger}$ Songyang Pu,$^{2}$ M. A. Mueed,$^{1}$ \\M. K. Ma,$^{1}$ K. A. Villegas Rosales,$^{1}$ Y. J. Chung,$^{1}$ L. N. Pfeiffer,$^{1}$ \\K. W. West,$^{1}$ K. W. Baldwin,$^{1}$ J. K. Jain,$^{2}$ M. Shayegan$^{1, *}$\\
\\
\normalsize{$^{1}$Department of Electrical Engineering, Princeton University,}
 \\\normalsize{Princeton, New Jersey 08544, USA}
\\
\normalsize{$^{2}$Department of Physics, 104 Davey Lab, Pennsylvania State University,}
 \\\normalsize{University Park, Pennsylvania 16802, USA}
\\
 \\\normalsize{$^{\dagger}$These authors contributed equally to this work.}
 \\
\normalsize{$^\ast$To whom correspondence should be addressed; E-mail:  shayegan@princeton.edu.}
}

\date{}
\renewcommand{\figurename}{\textbf{Fig.}}

\begin{document} 


\baselineskip24pt


\maketitle 

\begin{sciabstract}
In 1929 Felix Bloch suggested that the paramagnetic Fermi sea of electrons should make a spontaneous transition to a fully-magnetized state at very low densities, because the exchange energy gained by aligning the spins exceeds the enhancement in the kinetic energy. We report here the observation of an abrupt, interaction-driven transition to full magnetization, highly reminiscent of Bloch ferromagnetism that has eluded experiments for the last ninety years. Our platform is the exotic two-dimensional Fermi sea of composite fermions at half-filling of the lowest Landau level. Via quantitative measurements of the Fermi wavevector, which provides a direct measure of the spin polarization, we observe a sudden transition from a partially-spin-polarized to a fully-spin-polarized ground state as we lower the composite fermions' density. Our detailed theoretical calculations provide a semi-quantitative account of this phenomenon.
\end{sciabstract}

The ground state of a dilute system of electrons has long been a topic of theoretical fascination, because it represents a prototypical system with strong correlations \cite{Bloch.1929, Wigner.1934, Stoner.1947, Attaccalite.2002}. At low densities the interaction energy dominates over the kinetic energy, thereby enhancing the importance of electron correlations. Wigner famously predicted an electron crystal at very low densities \cite{Wigner.1934}, the search for which has driven many exciting developments. In another seminal paper \cite{Bloch.1929} Bloch predicted that electrons should spontaneously magnetize at low densities, because the gain in exchange interaction energy due to the alignment of all spins outweighs the increase in the kinetic energy. It is convenient to characterize the system with the dimensionless parameter $r_s$, the average inter-electron distance in units of the effective Bohr radius (equivalently, $r_s$ is also the ratio of the Coulomb to Fermi energies). For a two-dimensional electron system (2DES), sophisticated quantum Monte Carlo calculations \cite{Attaccalite.2002} indicate that such a transition should occur when $r_s$ exceeds 26, followed by another transition to a Wigner crystal state for $r_s > 35$. For a 2DES, we have $r_s= (me^2/4 \pi \hbar^2 \varepsilon \varepsilon_0)/ (\pi n)^{1/2}$, where $m$ is the electron effective mass, $\varepsilon$ is the dielectric constant, and $n$ is the 2DES density. Achieving a very dilute 2DES with low disorder, however, is extremely challenging. For example, in a GaAs 2DES ($m=0.067 m_0$ where $m_0$ is the free electron mass and $\varepsilon=13$), $r_s\simeq26$ corresponds to a density of $n \simeq 4.6 \times 10^8$ cm$^{-2}$, which is indeed very difficult to attain \cite{Sajoto.PRB.1990, Zhu.PRL.2003}. In GaAs 2D \textit{hole} systems ($m \simeq 0.4 m_0$), large $r_s$ values can be reached more easily, and in fact hints of a Wigner crystal formation near $r_s \simeq 35$ were reported \cite{Yoon.PRL.1999}. However, partly because of the strong spin-orbit interaction, extraction of the spin susceptibility and polarization of 2D holes is not straightforward \cite{Winkler.PRB.2005}.

While the spin polarization of an interacting 2DES has always been of great interest, it was thrust into the limelight in the 1990s in the context of the enigmatic metal-insulator transition (MIT) in dilute 2D carrier systems \cite{Zhu.PRL.2003, Yoon.PRL.1999, Winkler.PRB.2005, Vakili.PRL.2004, Kravchenko.Rep.Prog.Phys.2004, Spivak.Rev.Mod.Phys.2010, Pudalov.PRB.2018, Li.PRB.2019}. Numerous experiments revealed that the spin/valley polarization in 2D carrier systems plays an important role in the temperature dependence of conductivity. There were even reports that the MIT is linked to the full spin-polarization transition (for reviews, see, e.g., \cite{ Kravchenko.Rep.Prog.Phys.2004, Spivak.Rev.Mod.Phys.2010}). However, in a nearly-ideal (single-valley, isotropic, very thin) 2DES confined to a narrow AlAs quantum well, it was observed that the spin susceptibility closely follows the Monte Carlo calculations' results up to the highest experimentally achieved $r_s$ ($\simeq10$) and, importantly, remains finite as the 2DES goes through the MIT at $r_s \simeq 8$ \cite{Vakili.PRL.2004}. It is fair to say that an experimental observation of the transition to full spin polarization as manifested, e.g., by a divergence of the spin susceptibility, or a sudden increase in the Fermi wavevector, has been elusive so far (see, e.g. \cite{Pudalov.PRB.2018, Li.PRB.2019, Vermeyen.PRA.2018}).

We report here the observation of a sudden, Bloch-type, interaction-driven transition to a fully-magnetized ground state in a 2D system of composite fermions (CFs) near Landau level filling factor $\nu=1/2$ as their density is reduced. These exotic, fermionic quasiparticles form in 2DESs exposed to a perpendicular magnetic field ($B$), and each is composed of two magnetic flux quanta and an electron \cite{Jain.PRL.1989, Halperin.PRB.1993, Jain.2007}. Although of a collective origin, CFs behave like ordinary fermions in many respects. At $\nu=1/2$, having absorbed two flux quanta, the CFs act as if there is no external magnetic field, and occupy a metallic Fermi sea state with a well-defined Fermi wavevector \cite{Halperin.PRB.1993, Jain.2007}. Their Fermi sea and cyclotron orbits at small effective magnetic fields ($B^* = B-B_{\nu=1/2}$), and quantized energy levels at larger $B^*$, have been observed experimentally \cite{Jain.2007}. In our study, we probe the CF Fermi sea via geometric resonance (GR) measurements. The key idea is that when a weak, one-dimensional periodic perturbation is applied to the 2DES, if the CFs can complete a cyclotron orbit without scattering, then they exhibit a GR when their orbit diameter equals the period of the perturbation [Fig. 1a]. Such a resonance provides a direct and quantitative measure of the CFs' Fermi wavevector \cite{Halperin.PRB.1993, Jain.2007, Willett.PRL.1993, Kang.PRL.1993, Willett.PRL.1999, Smet.PRL.1999, Kamburov.PRL.2014}. 

\begin{figure*}[t!]
\includegraphics[width=1\textwidth]{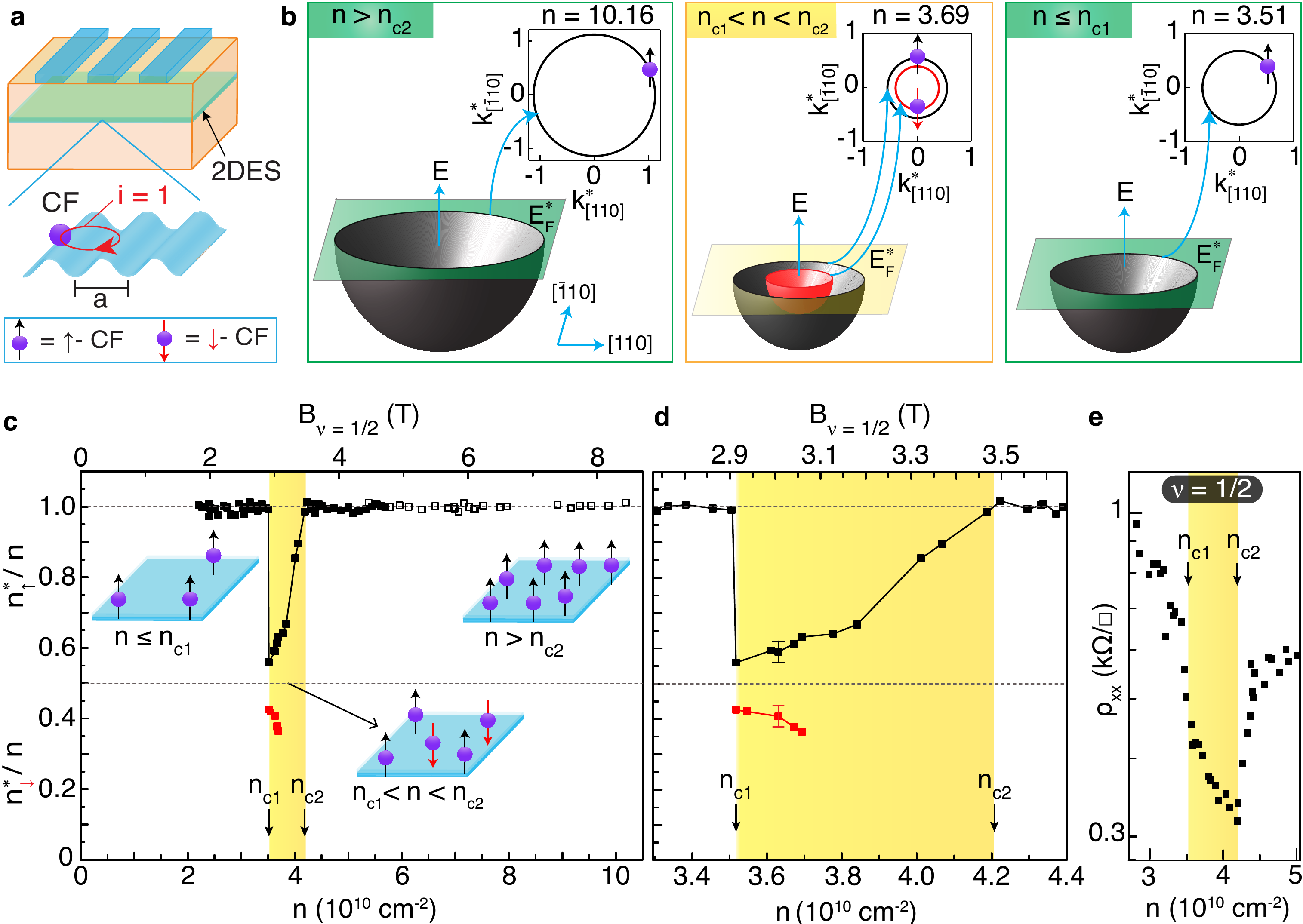}
\caption{\label{fig:Fig1} Evolution of the spin polarization of CFs as a function of the electron density ($n$). \textbf{a}, Our experimental technique to probe the density of CFs: Lateral surface superlattice of period $a$, inducing a periodic density perturbation in the 2DES. When the cyclotron orbit of the CFs becomes commensurate with the period of the perturbation, the $i = 1$ GR occurs. \textbf{b}, Schematics of CF energy vs wavevector. Cuts at the CF Fermi level ($E^*_F$) lead to circular Fermi contours as shown in the insets; the units for $k_{[110]}$ and $k_{[\bar{1}10]}$ are $10^{8}$ m$^{-1}$. These figures capture the evolution of the CF Fermi sea from fully magnetized to partially magnetized and back to fully magnetized as we lower the density. Black and red contours denote the $\uparrow$-spin and $\downarrow$-spin CFs, respectively. The areas encircled by these Fermi contours yield the densities $n^*_{\uparrow}$ and $n^*_{\downarrow}$ of CFs with majority and minority spins, respectively. \textbf{c}, The measured $n^*_{\uparrow}$ and $n^*_{\downarrow}$, normalized to $n$, plotted against $n$. At large $n$, the CFs are fully spin polarized, i.e., $n^*_{\uparrow}/n\simeq 1$. As the density is lowered below $n_{c2} \simeq 4.2$, the full spin polarization is lost and $n^*_{\uparrow}/n<1$. Remarkably, at even lower densities  $n \leq n_{c1} = 3.51$, the CFs experience an itinerant transition to a fully-spin-polarized state. Filled and open squares represent data from two samples from different wafers. \textbf{d}, A zoomed-in view of the yellow band ($n_{c1}<n<n_{c2}$) where the CFs are partially spin polarized. Typical error bars ($\pm4\%$) are also shown. \textbf{e}, Resistivity ($\rho_{xx}$) at $\nu=1/2$ vs $n$ for CFs. The pronounced minimum in $\rho_{xx}$ in the yellow band signals that the CFs are partially spin polarized in this density range; see SM \cite{SM} for details.}
\end{figure*}  

Our samples are 2DESs confined to modulation-doped, GaAs/AlGaAs heterostructures, grown by molecular beam epitaxy. The 2DESs have densities ranging from $10.16$ to $2.20$ in units of $10^{10}$ cm$^{-2}$ which we use throughout this paper. In this density range $r_s$ is $1.7-3.7$; see methods for more details. We use a back gate to tune the density. In our GR measurements, we impose a small periodic density modulation, the estimated magnitude of which is less than $\simeq0.5\%$; see methods. As illustrated in Fig. 1a, this is achieved by fabricating a one-dimensional, strain-inducing superlattice with period $a\simeq200$ nm on the surface of a lithographically-defined Hall bar \cite{ Kamburov.PRL.2014}.

Figures 1b--e highlight our key experimental finding. They show the evolution of the CF Fermi sea and measured densities of CFs with majority ($n^*_{\uparrow}$) and minority ($n^*_{\downarrow}$) spin, normalized by the total electron density $n$, as $n$ is varied. (Throughout the paper, we denote CF parameters by an ``$^{*}$"). At high densities ($n>n_{c2}= 4.2$), we find that CFs are fully spin polarized, i.e., $n^*_{\uparrow}= n$. This is consistent with previous reports \cite{Jain.2007, Willett.PRL.1993, Kang.PRL.1993, Kamburov.PRL.2014, Willett.PRL.1999, Smet.PRL.1999, Dementyev.PRL.1999, Melinte.PRL.2000, Tracy.PRL.2007, Liu.PRB.2014, Park.PRL.1998}. As is well documented experimentally and theoretically (see, e.g., \cite{Jain.2007, Liu.PRB.2014, Park.PRL.1998}), at high densities, or equivalently high $B$, when the ratio ($\alpha$) of the Zeeman energy ($E_Z=g\mu_B B$) to Coulomb energy [$E_C=e^2/4\pi\varepsilon \varepsilon_0 l_B$; $l_B=(\hbar/eB)^{1/2}$] exceeds a critical value ($\alpha_c \sim 0.01$), the CFs are fully spin polarized. In our samples, $n_{c2}=4.2$ corresponds to $\alpha \simeq 0.01$, in very good agreement with what is expected. As we lower $n$ below $4.2$, CFs lose their full spin polarization; this is also in accordance with previous measurements. Now, as the density is lowered even further, if the CFs are assumed to be non-interacting, one would expect spin polarization to continue to decrease. There is no reason for the CFs to become fully spin polarized again since $\alpha =E_Z/ E_C$ is further decreased. Remarkably, however, as the density is lowered to $n_{c1} = 3.51$, CFs make a sudden transition back to a fully-magnetized state. At even lower densities, the CFs remain fully magnetized. We attribute this transition to a Bloch-type, interaction-driven transition to a ferromagnetic state.

Next we describe how we deduce the degree of CFs' spin polarization from GR measurements. When the CFs experience a small $B^*$, they orbit in a circular cyclotron motion with a radius of $R_{c}^{*} = \hbar k_F^{*}/eB^{*}$, the size of which is determined by the magnitude of the CFs' Fermi wavevector, $k_F^{*}$. If the CFs have a sufficiently long mean-free-path so they can complete a cyclotron orbit ballistically, then a GR occurs when the orbit diameter becomes commensurate with the period ($a$) of the perturbation [Fig. 1a]. Quantitatively \cite{Willett.PRL.1999, Smet.PRL.1999, Kamburov.PRL.2014}, when $2R_{c}^{*}/a=i+1/4$ ($i=1,2,3,...$), GRs manifest as minima in magneto-resistance at $B_{i}^{*}=2\hbar k_F^{*}/ea(i+1/4)$. Thus, $k_F^{*}$ can be deduced directly from the positions of $B_{i}^{*}$. Using the measured $k_F^{*}$, we can extract $n^{*}_{\uparrow}$ and $n^*_{\downarrow}$ from the relation $k_{F}^* = (4\pi n^*_{\uparrow, \downarrow})^{1/2}$. 

\begin{figure*}[t!]
\includegraphics[width=1\textwidth]{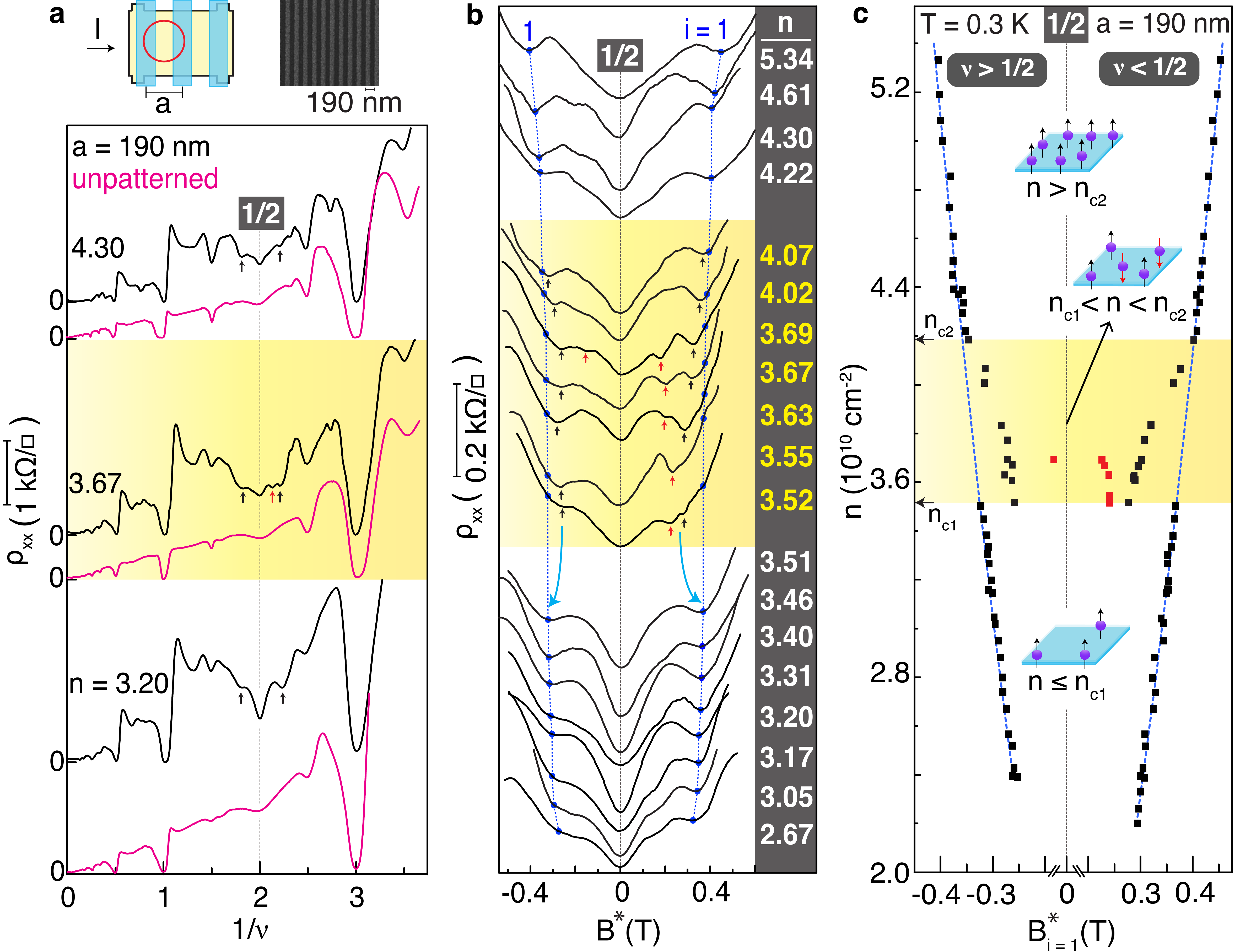}
\caption{\label{fig:Fig2} GR features of CFs near $\nu = 1/2$. Magneto-transport traces, all taken at $T=0.30$ K, are shown. In each panel the traces are vertically offset for clarity and the 2DES densities ($n$), in units of $10^{10}$ cm$^{-2}$, are given for each trace. \textbf{a}, Data plotted against $1/\nu$ for three densities, one in each of the three ($n > n_{c2}$,  $n_{c1} < n < n_{c2}$, and $n < n_{c1}$) regions, showing well-developed fractional quantum Hall states. Traces taken from the section patterned with an $a=190$ superlattice exhibit CF GR features near $\nu = 1/2$ demonstrated by the resistance minima flanking $\nu = 1/2$ (marked by arrows). The unpatterned section (magenta) traces show no such features. The upper panel shows a schematic of the Hall bar and a representative scanning electron micrograph of the superlattice used in our measurements. \textbf{b}, Evolution of CF GR features near $\nu = 1/2$ with density. Blue dots mark the $expected$ positions for the $i = 1$ GR for fully-spin-polarized CFs. The yellow band in all three panels indicates the $n_{c1} < n < n_{c2}$ region, where the CFs are only partially spin polarized, and the location of the CF GR minima (marked by vertical arrows) deviate from the blue dots. \textbf{c}, $B^*_{i=1}$ of CF GR plotted against $n$. The dotted blue curves show the \textit{expected} $B^*_{i=1}$ as a function of $n$ for fully-spin-polarized CFs. Black and red squares denote the CF GR features corresponding to the two $\uparrow$ and $\downarrow$ spin species, respectively. }
\end{figure*}

In Fig. 2 we focus on the densities close to $n_{c1}$ and $n_{c2}$ (see Figs. 1c,d), and show representative magneto-resistance traces each exhibiting well-developed GR features flanking a deep, V-shaped minimum at $\nu = 1/2$. The traces in Fig. 2a, which show magneto-resistance over a large range of $\nu$, attest to the high sample quality as evinced by the presence of fractional quantum Hall states, such as $\nu = 1/3$ and $2/3$, even at very low densities. In Fig. 2b we zoom in very close to $\nu = 1/2$, and show the traces as a function of the effective magnetic field seen by the CFs, $B^*=B-B_{\nu-1/2}$. The GR minima are clearly seen in these traces at $|B^*| \simeq 0.2$ to $0.5$ T. In each trace, we mark the \textit{expected} field positions of the $i = 1$ CF GR (blue dots) assuming a fully-spin-polarized CF Fermi sea. Note that, when CFs are fully spin polarized, their density was found experimentally to be the minority carrier density (electrons for $\nu<1/2$ and holes for $\nu>1/2$) \cite{Kamburov.PRL.2014}. The expected positions marked by the blue dots in Fig. 2b traces are based on this assumption.

\begin{figure*}[t!]
\includegraphics[width=1\textwidth]{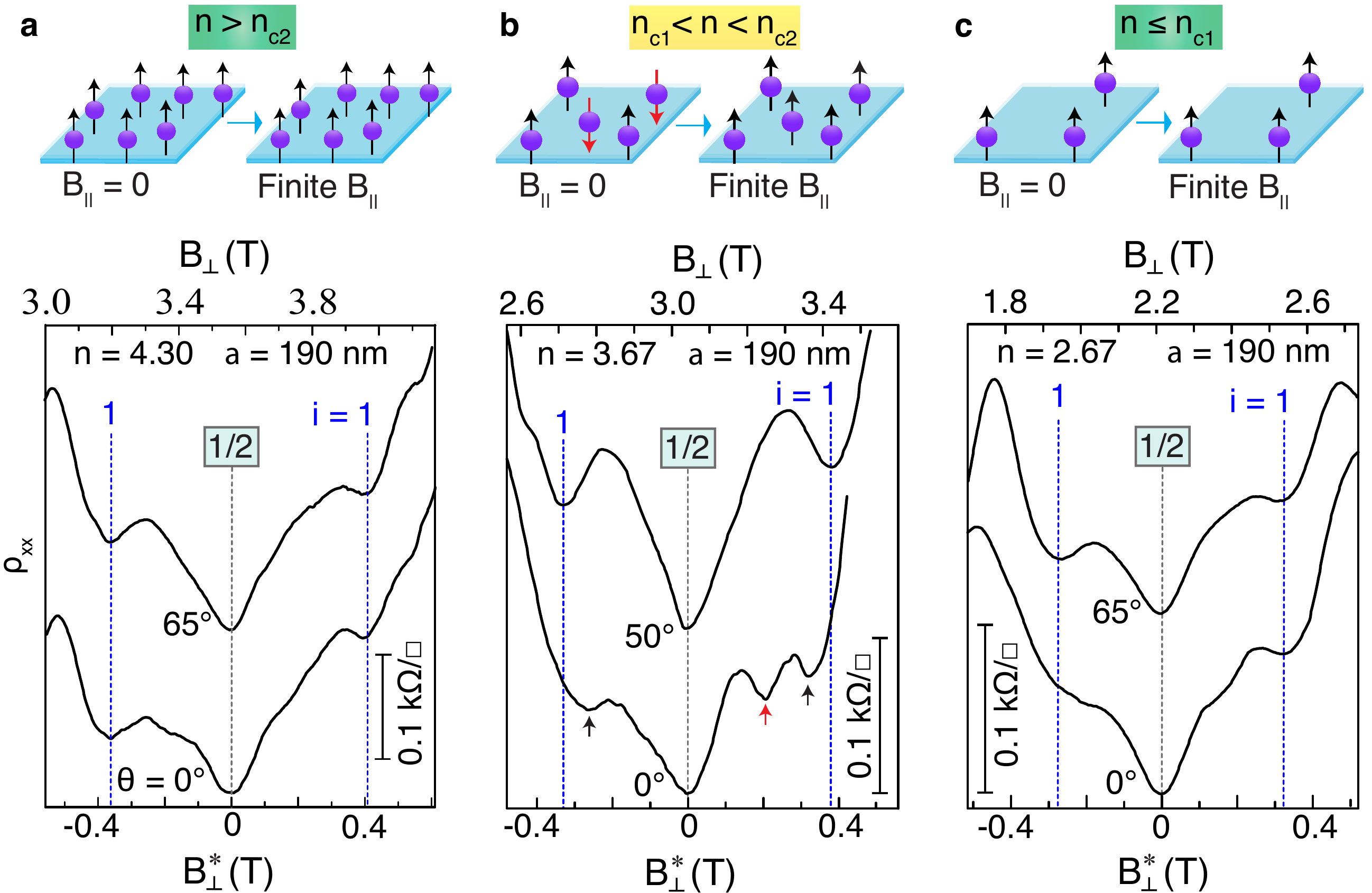}
\caption{\label{fig:FigS1} Tilt evolution of the CF GR features near $\nu = 1/2$ for: \textbf{a}, $n = 4.30$, \textbf{b}, $n = 3.67$, and \textbf{c}, $n = 2.67$. Traces are vertically offset for clarity; the tilt angle $\theta$ is given for each trace. The $expected$ positions for the $i = 1$ GRs for fully-magnetized CFs are marked with vertical blue lines. In all three panels, the scale for the applied external field $B_{\perp}$ is shown on top while the bottom scale is the effective magnetic field $B^*_{\perp}$ experienced by the CFs. At $n = 3.67$ (\textbf{b}), CFs near $\nu=1/2$ are partially spin polarized as evinced from the CF GR minima marked by arrows. However, in the presence of a sufficiently large $B_{||}$ (see the $\theta=50^o$ trace), the two minima on the $B^*>0$ side disappear and are replaced by a single minimum whose position aligns with the expected position for fully-spin-polarized CFs. The position of the GR minimum on the $B^*<0$ side also moves to the expected fully-polarized position. These traits are consistent with the presence of a partially polarized state for $n_{c1} < n < n_{c2}$ when $B_{||} = 0$ ($\theta=0^O$). On the other hand,  when the CFs are fully spin polarized, namely when $n > n_{c2}$ (\textbf{a}) or $n \leq n_{c1}$ (\textbf{c}), the locations of the CF GR minima remain unchanged as a function of $B_{||}$, and are consistent with the CFs being fully polarized for $B_{||}= 0$ and $B_{||}> 0$. See SM \cite{SM} for more details. The cartoons above the panels elucidate the effect of $B_{||}$. }
\end{figure*}

Starting from the highest density [uppermost trace in Fig. 2b], we note that the observed minima match the expected positions very well. This trend is seen in all the traces in the range $n>n_{c2}$. At lower densities, when $n_{c1}<n<n_{c2}$ (yellow band in Fig. 2b), however, the observed positions of the GR minima deviate from the $B_{i=1}^*$ expected for fully-spin-polarized CFs and move closer to $\nu=1/2$. Remarkably, at $n = 3.69$ and slightly lower densities, we observe two $i=1$ GR minima on the $B^*>0$ side of $\nu=1/2$ (Fig. 2b). We associate these two minima with the $i=1$ GR of CFs with different spins, and plot the deduced $n^{*}_{\uparrow}$ and $n^{*}_{\downarrow}$ in the yellow bands of Figs. 1c,d. Note that the sum of $n_{\uparrow}^*$ and $n_{\downarrow}^*$ equals the total density, as expected. (Note also that the double minima are observed primarily on the $B^*>0$ side of $\nu=1/2$; we do not know the reason for this asymmetry.) As detailed in the SM \cite{SM}, we have additional evidence for the partial spin polarization of CFs in the range $n_{c1}<n<n_{c2}$. The evidence includes GR data taken in tilted magnetic fields (see Fig. 3 and also Fig. S3), and also the observation of a pronounced minimum in the resistivity of the CFs at $\nu=1/2$ in the density range $n_{c1}<n<n_{c2}$ (see Figs. 1e and S2). The minimum is consistent with the enhanced screening of the disorder potential when the CFs are partially spin polarized \cite{SM}.

The most striking finding of our data is captured below the yellow band in Fig. 2b. When the density is lowered below 3.52, suddenly the traces near $\nu=1/2$ become simple again and, most notably, the positions of the observed GR minima on the two sides of $\nu=1/2$ agree very well with the expected $B^{*}_{i=1}$ for fully-spin-polarized CFs (blue dots). This observation provides direct and quantitative evidence that CFs at very low densities make a sudden transition to a fully-magnetized state. We summarize the measured $B^{*}_{i=1}$ as a function of $n$ in Fig. 2c. Figure 2c illustrates that starting from the highest density and down to $n_{c2}$, $B^{*}_{i=1}$ follows the blue curves which denote the \textit{expected} $B^*_{i=1}$ for full spin polarization. For $n_{c1}<n<n_{c2}$, $B^{*}_{i=1}$ deviates gradually from the blue curves (see the yellow band), indicating a loss of magnetization. Finally, below $n_{c1}$, $B^{*}_{i=1}$ suddenly aligns with the blue curves, signaling a Bloch-like transition to full spin polarization.

The surprising experimental results described in Figs. 1--3 prompted us to perform a detailed theoretical investigation of the spin physics of the CF liquid at $\nu=1/2$ as a function of density.   
If one neglects the effect of Landau level mixing (LLM) and considers an ideal 2DES with zero layer thickness, then, both the inter-CF interaction and the CF kinetic energies are proportional to the Coulomb energy, which is the only scale in the problem. In this case, the transition from a fully-spin-polarized state to a partially-polarized-state occurs at a density-independent value of $E_Z/E_C$, and, as seen below, no Bloch transition occurs at $\nu=1/2$ as a function of density. The transition seen at $\nu=1/2$ is therefore likely driven by LLM, which alters the residual interaction between CFs in a complicated manner. The parameter $r_s$ characterizes the strength of LLM; for $\nu=1/2$ we have $r_s=2\kappa$, where $\kappa=E_C/\hbar\omega_c$ is the standard LLM parameter.

\begin{figure*}[t!]
\includegraphics[width=.7\textwidth]{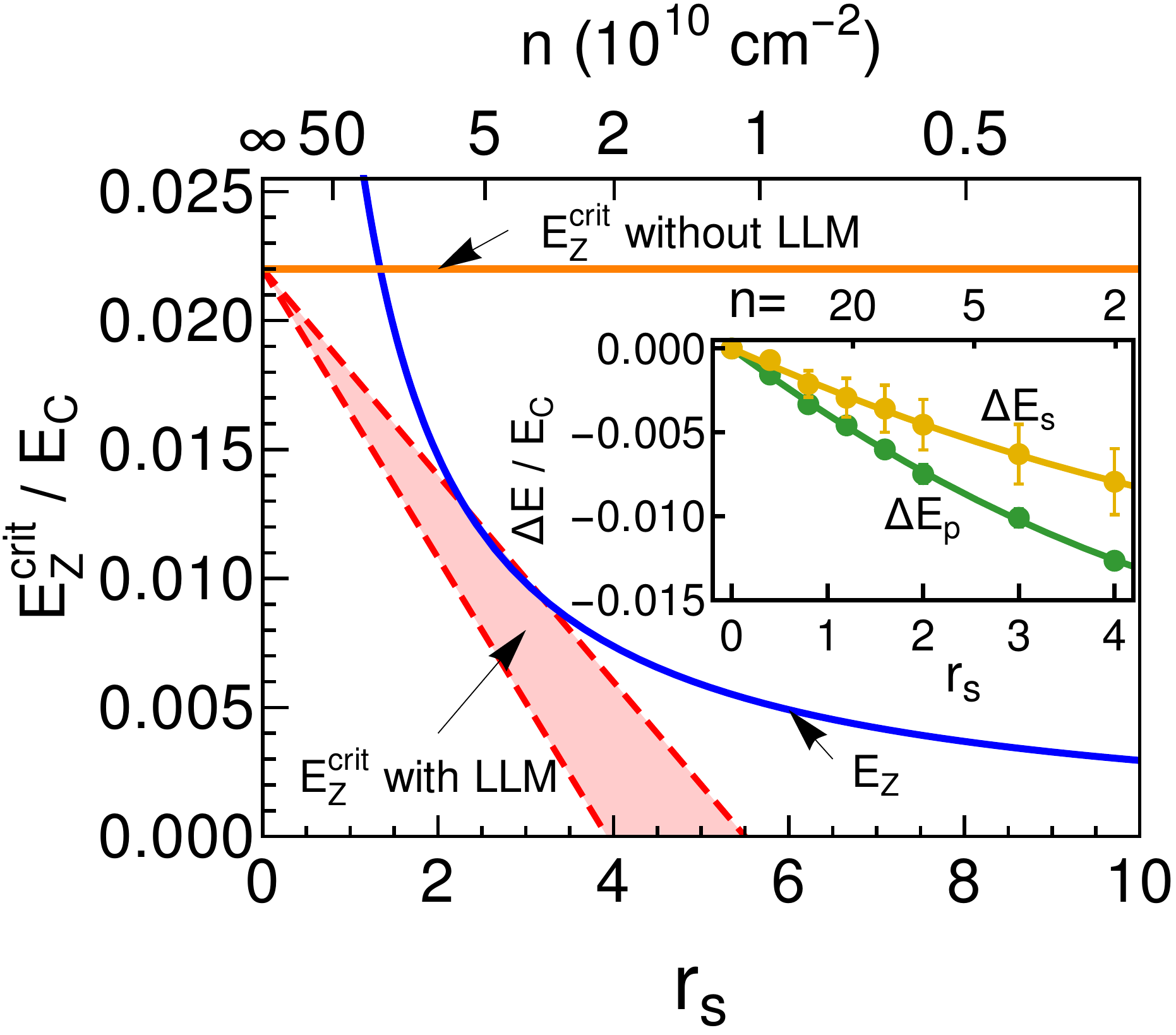}
\caption{\label{fig:theory} The critical Zeeman energy $E_Z^{\rm crit}$ as a function of $r_s$ or the density $n$ (shown on top), for the fixed filling factor $\nu=1/2$. [The two are related as $n\approx (32.8 /r_s^2) \times10^{10}$ cm$^{-2}$ for parameters appropriate for GaAs 2DESs.] Note that all plotted energies are normalized to the Coulomb energy ($E_C$). The horizontal orange line shows $E_Z^{\rm crit}$ without LL mixing (LLM), whereas $E_Z^{\rm crit}$ with LLM lies in the red region (indicating the statistical uncertainty in our DMC calculation). The system is predicted to be fully spin polarized when the Zeeman energy $E_Z$ (blue line) exceeds $E_Z^{\rm crit}$. 
These results admit the possibility of two phase transitions in the vicinity of $r_s\approx 2.8$ or $n\approx 4.2\times 10^{10}$ cm$^{-2}$. 
The inset shows $\Delta E_p(r_s)=E_p(r_s)-E_p(r_s=0)$ and $\Delta E_s=E_s(r_s)-E_s(r_s=0)$ respectively, i.e. changes in the energies (per particle) of the fully-polarized and spin-singlet states as a function of LLM, obtained by extrapolating the energies of finite-size systems to the thermodynamic limit~\cite{SM}; the calculated behavior, $\Delta E_{p}(r_s)-\Delta E_s(r_s)\approx -\gamma r_s E_C$ with $\gamma=0.0012(2)$, is used for determining $E_Z^{\rm crit}$ as explained in the text. 
}
\end{figure*}

We incorporate LLM through the non-perturbative, fixed-phase, diffusion Monte Carlo (DMC) method \cite{Ortiz.PRL.1993}. (Wee SM \cite{SM} for details. This method has been employed previously to study the role of LLM on spin polarization of the fractional quantum Hall states \cite{Zhang.PRL.2016}, and the 
stability of the crystal state at low fillings \cite{Ortiz.PRL.1993,Zhao.PRL.2018}.)
We assume zero layer thickness -- this should be a good first-order approximation because the transverse wavefunction in the heterojunction geometry is narrow, and also because corrections due to LLM are more important than finite-width corrections for the parameters of interest. The calculated changes \cite{SM} in the  energies per particle for the fully-spin-polarized and spin-singlet CF Fermi seas, $\Delta E_p(r_s)=E_p(r_s)-E_p(0)$ and $\Delta E_s(r_s)=E_s(r_s)-E_s(0)$ respectively, are shown in the inset of Fig.~\ref{fig:theory}. It is notable that $E_p$ decreases faster than $E_s$ with increasing $r_s$, which is what causes the Bloch transition.

A direct comparison with experiment requires a consideration of $E_Z$. In the Stoner model of itinerant ferromagnetism~\cite{Stoner.1947}, which considers a contact interaction between oppositely-spin-polarized CFs, the fully-spin-polarized state is predicted to occur \cite{SM} for $E_Z\geq E_Z^{\rm crit}\equiv 4(E_p-E_s)$, where $E_Z^{\rm crit}$ is the critical Zeeman energy. From our DMC calculation (inset of Fig.~\ref{fig:theory}) we estimate $E_p(r_s)-E_s(r_s)=E_p(0)-E_s(0)-\gamma r_s E_C$ with $\gamma=0.0012(2)$, which gives $E_Z^{\rm crit}(r_s)=E_Z^{\rm crit}(0)-4\gamma r_s E_C$. With $E_Z^{\rm crit}(0)=0.022 E_C$ taken from an earlier calculation~\cite{Park.PRL.1998}, the resulting $E_Z^{\rm crit}(r_s)$  lies in the shaded red region in Fig.~\ref{fig:theory}. A phase transition occurs when $E_Z$, shown by the solid blue line, crosses $E_Z^{\rm crit}$. Our calculation admits the possibility of two transitions in the vicinity of $r_s\approx 2.8$, which corresponds to $n\approx 4.2\times 10^{10}$ cm$^{-2}$ for parameters of GaAs.  Given the various approximations and assumptions made in our model and the microscopic DMC calculations, and given that $E_Z^{\rm crit}$ depends sensitively on very small energy differences (which are merely a fraction of 1\% of the energies of the individual states), our theory should be considered only semi-quantitative, but it brings out the qualitative physics underlying the Bloch transition of CFs. The SM shows that transitions at the experimentally observed densities $n_{c1}$ and $n_{c2}$ can be obtained by fitting the values of $\gamma$ and $E^{\rm crit}_Z(r_s=0)$, which are in reasonable proximity to those obtained from the microscopic calculation. 

For $E_Z=0$, which can in principle be achieved by application of hydrostatic pressure, theory predicts a fully-spin-polarized state for $r_s > 4.6(6)$ (see Fig.~\ref{fig:theory}). One may ask why the Bloch transition for CFs should occur at a relatively small value of $r_s$ compared to $r_s\approx 26$ for electrons. The value of $r_s$ where the Bloch transition occurs for CFs is determined by a complex interplay between the inter-CF interaction, a remnant of the Coulomb interaction between electrons, and the CF kinetic energy. Intuitively, one factor that reduces the critical $r_s$ for CFs is their large mass (compared to the electron band mass), which diminishes the importance of the CF kinetic energy. 

In summary, we report an experimental demonstration of the elusive Bloch ferromagnetism. We observe this phenomenon in a system of unusual suspects, namely flux-electron CFs near LL filling $\nu=1/2$. The transition to a fully-magnetized phase happens abruptly at very low densities. Theoretical calculations demonstrate that the transition occurs due to enhanced LLM at very low CF densities, which favors a fully-magnetized ground state. Our findings highlight a hitherto unexpected result of inter-CF interaction for $\nu=1/2$. Meanwhile, the quest for the experimental observation of Bloch ferromagnetism in \textit{zero-field} interacting fermions continues.

\setcounter{figure}{0}
\renewcommand{\thefigure}{M\arabic{figure}}%
\renewcommand{\figurename}{\textbf{Fig.}}
\setcounter{equation}{0}
\renewcommand{\theequation}{M\arabic{equation}}

\section*{Methods}

Our samples are two-dimensional electron systems (2DESs) confined to modulation-doped, GaAs/AlGaAs heterostructures, grown by molecular beam epitaxy. The 2DESs are buried 190 nm underneath the surface to ensure good sample quality. They have densities ($n$) ranging from $10.16$ to $2.20$ in units of $10^{10}$ cm$^{-2}$ which we use throughout. The low-temperature mobilities ($\mu$) of our samples are in the range $1.1 - 0.5$ in units of $10^{7}$ cm$^2/$Vs which are very high and therefore favorable to ballistic transport of composite fermions (CFs). This is a necessary condition for observing the geometric resonance (GR) features. For our measurements, Hall bar samples were fabricated using standard photolithography techniques, and alloyed InSn contacts were used to contact the 2DESs. The samples comprise multiple Hall bar sections, each of length 100 $\mu$m and width 50 $\mu$m. Via electron-beam lithography and using a calixarene-based negative electron-beam resist, we patterned the sample surface with strain-inducing superlattices perpendicular to the current direction. Thanks to the piezoelectric effect in GaAs, the periodic strain from this surface superlattice propagates to the 2DES and leads to a small density modulation. The availability of multiple sections enables us to study different superlattice periods in the same sample. The different sections are patterned with periods ranging from 190 to 225 nm. Lastly, we fit each sample with an In back gates to tune the electron density.

The measurements were carried out in a $^3$He cryostat with a base temperature of 0.3 K via passing current ($I = 10$ nA, 13 Hz) perpendicular to the density modulation. Note that, we injected a very low measurement current to avoid polarizing the nuclear spins which might introduce an effective nuclear field to our 2DES thus affecting the electron Zeeman splitting \cite{Tracy.PRL.2007}. We also checked the up and down magnetic field sweeps and did not observe any noticeable hysteresis. Such absence of hysteresis is indeed expected from a system of CFs. In that context, it is important to note that CFs, by necessity, require a finite magnetic field to emerge. When the CFs become fully spin polarized, it makes most sense that the direction of their aligned spins is determined by the direction of the applied magnetic field. Therefore, as we sweep the magnetic field about $\nu=1/2$ (or the position of the GRs), one would not expect magnetic domains or hysteresis. We also stress that the relevant magnetic field for the spin polarization of the CFs is the total applied magnetic field $B$ \cite{Jain.2007} (which is always non-zero) and not the effective field $B^*$ (that passes through zero as one goes from $\nu>1/2$ to $\nu<1/2$).

Next we discuss the periodic potential modulation which we impose on CFs for GR measurements. The strength of this modulation depends on the depth of the 2DES from the surface and the period of the superlattice. For periods shorter than the 2DES depth, the higher harmonics of the potential modulation are attenuated, resulting in a weaker modulation amplitude; see Ref. \cite{Mueedstripe.PRL.2016} and references therein. Following the same analysis as in Ref. \cite{Mueedstripe.PRL.2016}, we can obtain a reasonable estimate of the strength of this periodic modulation: We normalize the superlattice period to the depth of the 2DES (190 nm for our samples), and then compare it to the 2DES of 135-nm-depth used in Ref. \cite{Mueedstripe.PRL.2016}. The strength of our 190-nm-period modulation turns out to be less than 0.5$\%$.

We stress that the period of the superlattice required to observe clear GR features depends on the electron density and the depth of the 2DES. This is because the CFs are fragile and require a very gentle modulation to exhibit clear GR features \cite{Shafayat5/2.PRL.2018}. In our samples, the depth of the 2DES is fixed and equals 190 nm. On the other hand, we tune the density via applying voltages to the back gate. Therefore, in order to observe clear GR features in all the densities in our experiments, we use superlattices of different periods depending on the electron density. For densities down to $n\simeq7.05$, we observe clear GR features for $a=225$ and $200$ nm. For lower densities we use $a=200$ nm (for $n\simeq6.99$ to $5.34$) and $a=190$ nm (for $n=5.57$ to $2.26$) to ensure an appropriate modulation. 

We show in Fig. M1, the magneto-resistance traces for the sample patterned with $a=190$ nm for four different densities in the ranges $n > n_{c2}$, $n_{c1} < n < n_{c2}$, and $n \leq n_{c1}$, and compare them with the corresponding unpatterned (reference) section of the Hall bar. Although small, the impact of the periodic modulation created by the $a=190$ nm superlattice manifests in the magneto-resistance traces as a V-shaped minimum at $\nu=1/2$, flanked by minima on both sides of $\nu=1/2$ which signal the GR of CFs with the periodic potential (upper traces in Fig. M1). On the other hand, as seen from the lower traces in all four panels of Fig. M1, the unpatterned section exhibits no such features flanking $\nu=1/2$. We note that, the V-shaped minimum at $\nu=1/2$ in the $a=190$ nm traces, i.e. the positive magneto-resistance around half-filling, is attributed to open orbits of CFs, analogous to the open orbits of carriers in the classical magnetic breakdown mechanism; see Refs. \cite{Smet.PRL.1999, Kamburov.PRL.2014}. 

\begin{figure*}[t!]
\includegraphics[width=1\textwidth]{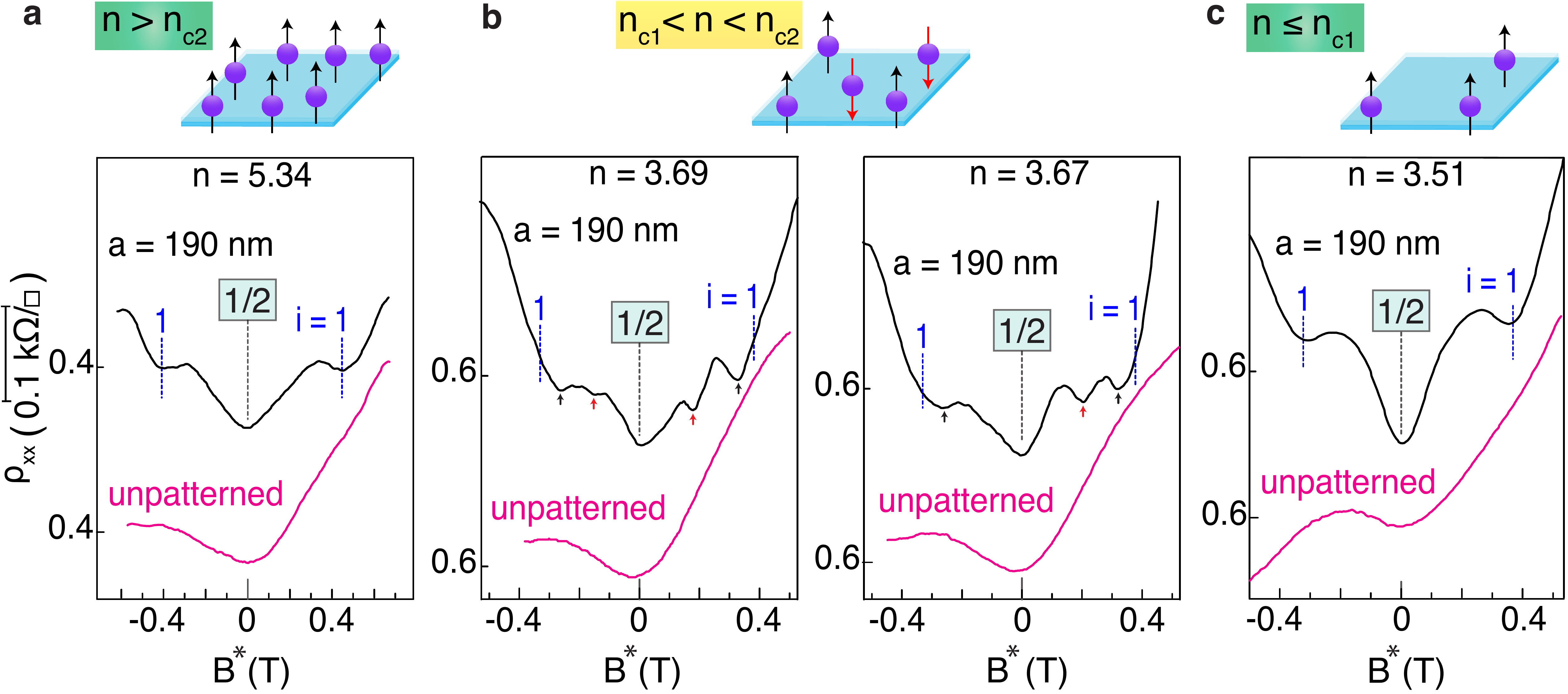}
\caption{\label{fig:FigM1} 
Comparison between the patterned and unpatterned sections of the Hall bar. \textbf{a-c}, Magneto-resistance traces for four densities, taken in the patterned ($a=190$ nm) and unpatterned sections, are shown. Traces are vertically offset for clarity. While we observe clear GR minima flanking $\nu=1/2$ in the traces taken from the section patterned with an $a=190$ superlattice, the unpatterned section traces do not exhibit any such features, as expected. The cartoons in the upper panels show the spin configurations for different densities.}
\end{figure*}

Finally, as seen in Fig. 2b, the traces show pronounced GR minima on the $\nu<1/2$ ($B^*>0$) side down to very low densities ($n \simeq 2.7$). Therefore, we can extract the location of the GR minima on the $\nu<1/2$ side with good accuracy. On the other hand, at densities below $n \simeq 2.7$, and especially on the  $\nu>1/2$ ($B^*<0$) side, the CF GR minima are less pronounced. Therefore, to accurately extract the $B_{i=1}^*$ for $n \lesssim 2.7$, we perform a simple background subtraction process. We emphasize that, this extraction process is done for densities $n \lesssim 2.7$ that are well below $n=n_{c1}$ where the Bloch transition occurs. Therefore, it has no bearing on our conclusions.

\section*{Data Availability Statement}

Data that support the plots within this paper and other findings of this study are available from the corresponding author upon reasonable request.

\section*{Acknowledgments}
We acknowledge support through the U.S. Department of Energy Basic Energy Science (Grant DEFG02-
00-ER45841) for measurements, and the National Science Foundation (Grants DMR 1709076, ECCS 1906253, and MRSEC DMR 1420541), and the Gordon and Betty Moore Foundation (Grant No. GBMF4420) for sample fabrication and characterization. The theoretical work at Penn State (T.Z., S.P. and J.K.J.) was supported in part by the U. S. Department of Energy, Office of Basic Energy Sciences, under Grant No. DE-SC0005042.  J.K.J. thanks the the Indian Institute Science, Bangalore, where part of this work was performed, for their hospitality, and the Infosys Foundation for making the visit possible. M.S. acknowledges a QuantEmX travel grant from the Institute for Complex Adaptive Matter (ICAM) and the Gordon and Betty Moore Foundation through Grant No. GBMF5305. We also thank R. Warburton and R. Winkler for illuminating discussions.   

\section*{Author contributions}
M. S. H. fabricated the devices, performed the measurements, and analyzed the data. M. S. H., M. A. M., J. K. J., and M. S. discussed the data. T. Z., S. P., and J. K. J. performed the theoretical calculations. Y. J. C., L. N. P., K. W. W., and K. W. B. grew the quantum well samples via molecular beam epitaxy. M. K. M. and K. A. V. R. helped with the measurements. M. S. H., T. Z., S. P., J. K. J., and M. S. co-wrote the manuscript with input from all co-authors.

\end{document}